\documentclass{emulateapj}

\slugcomment{Accepted for publication in ApJL}

\shorttitle{Specific Star Formation Rates to Redshift 5 from FDF and GOODS-S}
\shortauthors{Georg Feulner et al.}

\begin{document}


\title{Specific star formation rates to redshift 5 
from the FORS Deep Field \\and the GOODS-S Field\altaffilmark{1}}


\author{Georg Feulner\altaffilmark{2,3},
  Armin Gabasch\altaffilmark{3,4},
  Mara Salvato\altaffilmark{4},
  Niv Drory\altaffilmark{5},
  Ulrich Hopp\altaffilmark{3,4},
  Ralf Bender\altaffilmark{3,4}}


\altaffiltext{1}{Based on observations collected at the European
  Southern Observatory, Chile; proposal numbers 63.O-0005,
  64.O-0149, 64.O-0158, 64.O-0229, 64.P-0150, 65.O-0048, 65.O-0049,
  66.A-0547, 68.A-0013, and 69.A-0014. This work makes use of data
  obtained by the ESO GOODS/EIS project under Program ID LP168.A-0485.}
\altaffiltext{2}{Email: feulner@usm.lmu.de}
\altaffiltext{3}{Universit\"ats-Sternwarte M\"unchen, Scheinerstra\ss
  e 1, D--81679 M\"unchen, Germany}
\altaffiltext{4}{Max-Planck-Institut f\"ur Extraterrestrische Physik,
Giessenbachstra\ss e, D--85748 Garching bei M\"unchen, Germany}
\altaffiltext{5}{University of Texas at Austin, Austin, Texas 78712}


\begin{abstract}
We explore the build-up of stellar mass in galaxies over a wide
redshift range $0.4 < z < 5.0$ by studying the evolution of the
specific star formation rate (SSFR), defined as the star formation
rate per unit stellar mass, as a function of stellar mass and age. Our
work is based on a combined sample of $\sim 9000$ galaxies from the
FORS Deep Field and the GOODS-S field, providing high statistical
accuracy and relative insensitivity against cosmic variance. As at
lower redshifts, we find that lower-mass galaxies show higher SSFRs
than higher mass galaxies, although highly obscured galaxies remain
undetected in our sample. Furthermore, the highest mass galaxies
contain the oldest stellar populations at all redshifts, in principle
agreement with the existence of evolved, massive galaxies at $1 < z <
3$. It is remarkable, however, that this trend continues to very high
redshifts of $z \sim 4$. We also show that with increasing redshift
the SSFR for massive galaxies increases by a factor of $\sim 10$,
reaching the era of their formation at $z \sim 2$ and beyond. These
findings can be interpreted as evidence for an early epoch of star
formation in the most massive galaxies, and ongoing star-formation
activity in lower mass galaxies.
\end{abstract}





\section{Introduction}

In recent years, there has been considerable interest in the relation
of the stellar mass in galaxies and their star-formation rate (SFR),
since this allows to quantify the contribution of the recent star
formation to the build up of stellar mass for different galaxy masses.
\citet{Cowie1996} used $K$-band luminosities and [OII], H$\alpha$ or
ultraviolet (UV) fluxes to investigate this connection for a
$K$-selected sample of $\sim 400$ galaxies at $z<1.5$ and noted an
emerging population of massive, heavily star forming galaxies at
higher redshifts, a phenomenon they termed `downsizing'. Later on, the
`specific SFR' (SSFR), defined as the SFR per unit stellar mass, was
used to study this relation.

\citet{Guzman1997} derived the SSFR for 51 compact galaxies at $z <
1.4$ in the HDF flanking fields \citep{HDFN} finding no evidence for
an increase of the peak SSFR with redshift. \citet{BE00} studied 321
$I$-selected field galaxies at $z<1$ and detected a clear upper limit
on the SSFR moving to higher SFRs with increasing redshift. They
conclude that the most massive galaxies must have formed the bulk of
their stars before $z = 1$.  \citet{PerezGonzalez2003} and
\citet{Brinchmann2004} presented detailed investigations of the SSFR
in the local universe, while \citet{Fontana2003} used a deep
$K$-selected sample of $\sim 300$ galaxies in the HDF-S \citep{HDFSb}
to trace the SSFR to $z > 2$ and found more evidence for higher SSFRs
in the past, a result confirmed at $z < 1.5$ by \citet{Bauer2005}
using spectroscopic data for $\sim 350$ galaxies. \citet{Juneau2005}
found clear evidence for downsizing in a sample of $\sim 200$ galaxies
at $0.8 < z < 2$ from the GDDS, and \citet{munics7} used the MUNICS
data \citep{munics1, munics5} to study the SSFR of $\sim 6000$
galaxies with photometric redshifts to $z = 1.2$, placing strong
emphasis on the age of the stellar populations. They confirmed
previous results on the rise of the SSFR with redshift, but found in
addition that the highest mass galaxies are dominated by the oldest
stellar populations at all redshifts.

\citet{Hammer2005} obtained 15$\mu$m fluxes for $\sim 200$ $z>0.4$
galaxies and estimated that 15\% of all $M_B<-20$ galaxies are
luminous IR galaxies with SSFRs well above the range usually found
using other star-formation estimators. \citet{Bell2005} investigated
$\sim 1700$ $B$-selected galaxies at $z \simeq 0.7$ with photometric
redshifts, $\sim 25$\% of which could be detected at $24\mu$m. They
found that these galaxies typically have masses in the range $9.5
\lesssim M_\star \lesssim 11.0$ and SFRs of up to $\dot{\varrho}_\star
\lesssim 100 \: M_\odot \: \mathrm{yr}^{-1}$. \citet{Perez2005} used
about 8000 sources selected at 24$\mu$m to study the SSFR to $z\sim
3$, finding clear support for the downsizing picture.

\begin{figure*}
\begin{center}
\includegraphics[height=0.85\textwidth,angle=270]{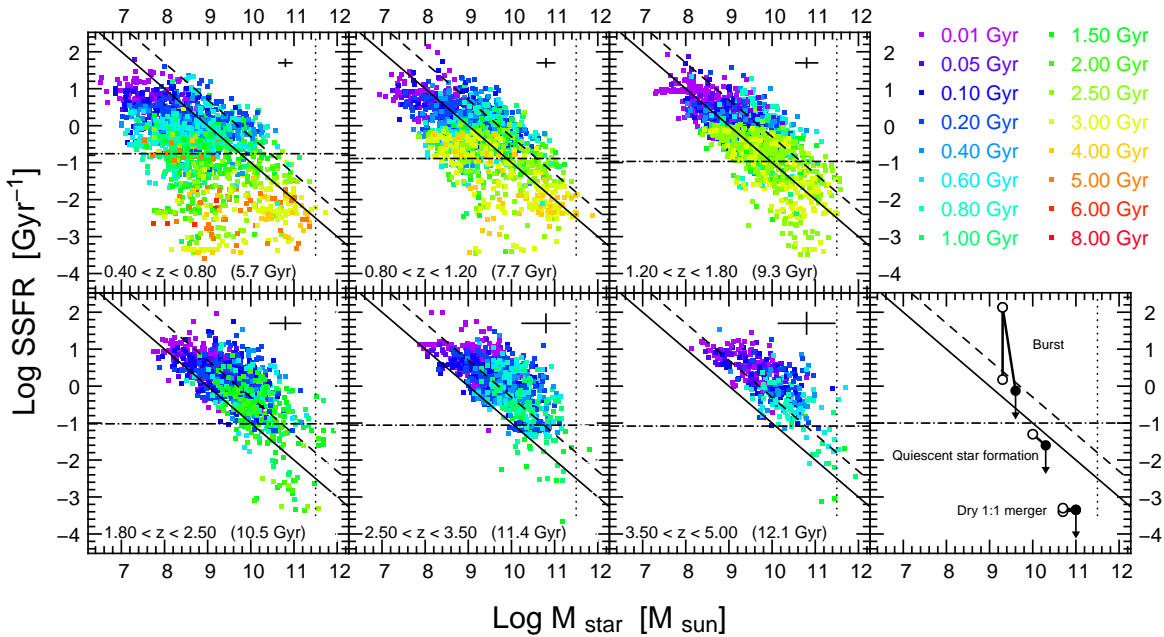}
\caption{\textit{Six panels to the left:} The SSFR as a function of
  stellar mass and redshift for the FDF and GOODS-S. Both the SFR and
  the mass are corrected for dust extinction. The solid and dashed
  lines correspond to SFRs of 1 $M_\odot \: \mathrm{yr}^{-1}$ and 5
  $M_\odot \: \mathrm{yr}^{-1}$, respectively. Objects are colored
  according to the age of the main component of the stellar population
  synthesis model fit to the photometry, ranging from 0.01~Gyr
  (purple) to 8~Gyr (red). The dot-dashed line is the SSFR required to
  double a galaxy's mass between each redshift epoch and today
  (assuming constant SFR); the corresponding look-back time is
  indicated in each panel. The error cross in each panel gives an idea
  of the typical errors, while the dotted line roughly represents the
  high-mass cut-off of the local stellar mass function \citep{munics6,
  Fontana2004, fdfmf}. \textit{Lower right-hand panel:} Examples for
  evolutionary paths yielding a doubling of a galaxy's mass, through
  quiescent star formation, through a burst of star formation
  superimposed on quiescent star formation, and through a dry
  equal-mass merger. Open symbols denote the starting point, filled
  symbols the final state; the doubling line is drawn for a lookback
  time of 10~Gyr ($z \simeq 2$). The arrows indicated the influence of
  gas consumption or loss.}
\label{f:ssfr}
\end{center}
\end{figure*}

In the following we present results on the SSFR of $\sim 9000$
galaxies at $0.4 < z < 5.0$ in the FORS Deep Field (FDF) and the
GOODS-S field, reaching higher redshifts than previous
investigations. This letter is organized as follows: We introduce the
galaxy sample and our method to derive the SSFR in
Section~\ref{s:samples}. In Section~\ref{s:ssfr} we present our
results on the evolution of the SSFR with
redshift. Section~\ref{s:dust} gives a brief account on the influence
of dust attenuation. In Section~\ref{s:paths} we discuss different
evolutionary paths in the SSFR--stellar mass diagram, before we
summarize our findings in Section~\ref{s:concl}.  Throughout we assume
$\Omega_m = 0.3$, $\Omega_\Lambda = 0.7$ and $H_0 = 70 \, \mathrm{km}
\, \mathrm{s}^{-1} \, \mathrm{Mpc}^{-1}$. All magnitudes are in the
Vega system.

\section{The Galaxy Samples}
\label{s:samples}

The FDF \citep{FDF1} offers photometry in the $U$, $B$, $g$, $R$, $I$,
834~nm, $z$, $J$ and $K$ bands and is complimented by deep
spectroscopic observations \citep{fdfspec}. In this letter we use the
$I$-selected sub-sample covering the deep central part of the field
($\sim$~40 arcmin$^2$) as described in \citet{fdflf1}, containing 5557
galaxies down to $I=26.4$.

Our $K$-band selected catalog for the GOODS-S field (Salvato et al.,
in prep.) is based on the publicly available 8 $2.5\times2.5$
arcmin$^2$ $J$, $H$, $Ks$ VLT/ISAAC images and contains 3297 galaxies
down to $K=23.5$ in $\sim$~50 arcmin$^2$.  The $U$ and $I$ images are
described in \citet{Arnouts2001}, the $B$, $V$, and $R$ images in
\citet{Schirmer2003}. The data were analyzed in a very similar way to
the FDF data and already used in \citet{fdfsfr} and
\citet{fdfmf}. Although both samples in itself are not large in area,
having two different lines of sight helps to overcome some of the
effects of cosmic variance. Furthermore, we chose our redshift
intervals large enough to further minimize the effect.

Photometric redshifts are derived using the method described in
\citet{photred}. We estimate the SFRs of our galaxies from the
spectral energy distribution by deriving the luminosity at $\lambda =
1500 \pm 100$\AA\ and converting it to an SFR as described in
\citet{Madau1998} assuming a Salpeter initial mass function
\citep{Salpe55}. Although this is an extrapolation for the lower
redshift bins, the results agree very well with our work at lower
redshifts \citep{munics7}. Stellar masses are computed from the
multi-color photometry using the same method as in \citet{fdfmf}. It
is described in detail and tested against spectroscopic and dynamical
mass estimates in Drory, Bender \& Hopp (\citeyear{masscal}). In
brief, we derive stellar masses by fitting a grid of stellar
population synthesis models by \citet{BC2003} with a range of star
formation histories (SFHs), ages, metallicities and dust attenuations
to the broad-band photometry. We describe the SFHs by a two-component
model consisting of a main component with a smooth SFH $\propto \exp
(-t/\tau)$ and a burst contributing up to 15\% in mass. We allow SFH
timescales $\tau \in [0.1,\infty]$~Gyr, metallicities $[\mathrm{Fe/H}]
\in [-0.6,0.3]$, ages between 0.01~Gyr and the age of the universe at
the objects' redshift, and independent extinction values of $A_V \in
[0.0,1.5]$ for the main component and the burst, respectively. We
adopt a Salpeter initial mass function for both components, with lower
and upper mass cutoffs of 0.1 and 100~$M_\odot$. The SFR is corrected
with the dust attenuation obtained for the burst component using the
extinction curve of \citet{Calzetti1997}. Note that at higher redshift
the uncertainty in the mass estimate increases since the observed $K$
band then probes the rest-frame blue or UV \citep{fdfmf}. We have
verified the uncertainty in the mass estimate by comparing masses at
lower redshifts from simulations with and without the NIR bands.

\section{The Specific Star Formation Rate}
\label{s:ssfr}

One way to explore the contribution of star formation to the growth of
stellar mass in galaxies of different mass is to study the redshift
evolution of the specific SFR (SSFR; \citealt{Guzman1997, BE00}) which
is defined as the SFR per unit stellar mass. In Figure~\ref{f:ssfr} we
present the SSFR as a function of stellar mass and age for six
different redshift bins covering the range $0.4 < z < 5.0$. We have
convinced ourselves that the distributions of galaxies from the FDF
and GOODS-S are in very good agreement.

Several effects can be observed in Fig.~\ref{f:ssfr}. The upper
cut-off of the SSFR running essentially parallel to lines of constant
SFR and shifting to higher SFRs with increasing redshift was already
noted in earlier work \citep{BE00, munics7, Bauer2005}. This trend
seems to continue to the highest redshifts probed by our sample: While
at $z \sim 0.6$ we find $\mathrm{SFR}_\mathrm{max} \simeq 5 \: M_\odot
\: \mathrm{yr}^{-1}$, galaxies reach as much as
$\mathrm{SFR}_\mathrm{max} \simeq 100 \: M_\odot \: \mathrm{yr}^{-1}$
at $z \sim 4$. Note that this upper envelope is partly due to a
selection effect: Heavily dust obscured star bursts cannot be detected
in our sample \citep[see, e.g.,][]{Hammer2005, Perez2005}, but our
conclusions still hold for galaxies not heavily affected by dust
extinction (see the discussion below). Furthermore, it is evident from
the distribution of ages in this diagram, that the most massive
galaxies contain the oldest stellar populations, as has been shown
already in \citet{munics7} and \citet{fdfmf}. This is in agreement
with the `downsizing' scenario \citep{Cowie1996}

\section{The role of dust}
\label{s:dust}

The influence of dust extinction on the determination of the SSFR is
two-fold. First, heavily dust enshrouded objects might escape
detection because too much of the optical light is absorbed. Secondly,
objects might be detected, but their SFR (and stellar mass) might be
underestimated because of the increasing dust extinction in the UV. We
try to correct for the second effect by including dust attenuation in
our model fitting, and correcting both the SFR and the stellar mass
accordingly, but the correction will likely be underestimated for
extremely dusty objects. It is more complicated, of course, to
overcome the first effect.

In principle both sources of uncertainty could be overcome with
observations in the thermal infrared (IR), where the radiation
absorbed by the dust component is re-emitted. Note however, that this
approach suffers from confusion and identification problems and also
involves uncertainties in the conversion of the observed IR flux to
the total IR flux, and in the unknown contribution of dust heating by
old stellar populations, see, e.g., the discussion in
\citet{Bell2005}.

Luminous IR galaxies (LIRGs) are a well known population of dusty
galaxies with very high SFRs and thus also SSFRs \citep[see,
e.g.,][]{Hammer2005, Perez2005}. They can be interpreted as galaxies
experiencing a brief episode of heavy star formation triggered by
mergers or gas infall. Due to their limited gas supply, these galaxies
would spend most of their time in a `normal' state with lower
SSFR. Therefore, although our survey misses dust enshrouded star
forming galaxies, they can be considered as intermittent stages in the
evolution of galaxies. In particular our results on the existence of
massive evolved galaxies even at high redshifts remain unaffected.

\begin{figure}
\begin{center}
\includegraphics[width=0.75\columnwidth]{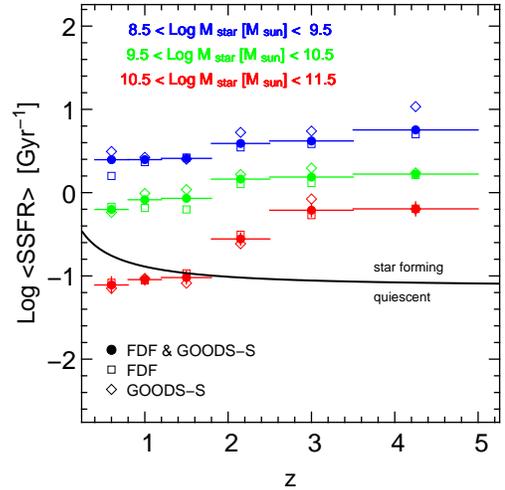}
\caption{Average SSFRs for galaxies with stellar masses of $\log
M_\star/M_\odot \in [8.5,9.5]$ (blue), $[9.5,10.5]$ (green) and
$[10.5,11.5]$ (red) and SFRs larger than $1 \: M_\odot \:
\mathrm{yr}^{-1}$ as a function of $z$ for FDF (open squares), GOODS-S
(open diamonds) and the combined sample (filled circles). The error
bar represents the error of the mean.  All numbers are given in
Table~\ref{t:meanssfrz}. The solid line indicates the doubling line of
Fig.~\ref{f:ssfr} which can be used to discriminate quiescent and
heavily star forming galaxies.}
\label{f:meanssfrz}
\end{center}
\end{figure}

\section{Evolutionary paths}
\label{s:paths}

In order to understand the role of various evolutionary paths in
Fig.~\ref{f:ssfr}, it is helpful to visualize schematically different
ways to double a galaxy's mass as shown in the lower right-hand panel
of Fig.~\ref{f:ssfr}. In the following we will discuss these paths in
more detail.

\textit{Quiescently star-forming galaxies:} A galaxy doubling its
stellar mass by quiescent star formation at $0.5 \: M_\odot \:
\mathrm{yr}^{-1}$ moves along a line of constant SFR towards the lower
right part of the diagram. Note that galaxies below the doubling line
in Fig.~\ref{f:ssfr} do not have enough time to double their mass
until today.

\textit{Starbursts:} In contrast to a quiescent galaxy, a starburst
can increase its mass in a shorter time interval, provided it has
enough gas to consume. Bursts of star formation may be triggered by
gas inflow or galaxy interactions, and quickly move a galaxy to high
SSFRs, where it stays for a brief period of time before it fades back
to `normal' SSFRs. Given the typical dusty nature of starburst
galaxies, they might escape detection in optical surveys during this
stage. However, since these bursts are typically brief, the galaxies
spend most of their time with the quiescent galaxies. Note that the
exact path depends on the details of the star formation history during
the burst phase, however, the only relevant parameter for the final
mass is the mass of the consumed gas.

\textit{Dry mergers:} Two galaxies undergoing a dry merger (i.e.\
without interaction-induced star formation) basically move to the
right in the diagram. We illustrate this with equal stellar mass
mergers; the stellar mass clearly doubles, while the final SSFR is the
average of the two initial SSFRs.

Note that the `true' endpoint of the galaxies' evolution will in all
three cases likely be lower than shown, since all three processes
diminish the limited gas supply. This is indicated by the down-ward
pointing arrows.

In the light of these evolutionary possibilities it is obvious that
the only two ways to form massive galaxies with old stellar
populations is by highly efficient early star formation in massive
haloes, or by dry merging of less massive galaxies harboring old
stars. Both scenarios can, in principle, be distinguished by analyzing
the redshift dependence of the SSFR in the most massive galaxies. The
result is presented in Fig.~\ref{f:meanssfrz}, where we show the
average SSFR as a function of redshift for galaxies in three mass
intervals. At redshifts $z \lesssim 2$, the most massive galaxies with
$\log M_\star/M_\odot \in [10.5,11.5]$ are in a quiescent state with
SSFRs not contributing significantly to their growth in stellar
mass. However, at redshifts $z \gtrsim 2$, the picture changes
dramatically: The SSFR for massive galaxies increases by a factor of
$\sim 10$ until we witness the epoch of their formation at $z \sim 2$
and beyond. The fact that we miss galaxies at $z \gtrsim 1$ which have
both high star-formation rate and mass, e.g.\ sub-mm galaxies at $z
\sim 2$ \citep[e.g.][]{Smail2002a}, might shift this formation epoch
to lower redshifts, but does not affect our conclusions.

\section{Discussion and Conclusions}
\label{s:concl}

It is remarkable that the most massive galaxies show evidence of
harboring the oldest stellar populations at all redshifts. Our sample
shows this effect robustly out to very high redshifts of $z \sim
4$. Note that this is in agreement with the findings of massive,
evolved galaxies in the population of Extremely Red Objects (EROs) at
$1 < z < 2$ \citep{tesis1, Cimatti2004, tesis3, tesis4} and among the
Distant Red Galaxies (DRGs) at $2 < z < 3$ \citep{Labbe2005}.
Apparently this trend continues to even higher redshifts, indicating a
very early formation epoch for the most massive galaxies in the
universe, favoring the `downsizing' picture \citep{Cowie1996}.

\begin{deluxetable}{crrr}
\tablecaption{Average SSFRs of galaxies as a function of $z$\label{t:meanssfrz}}
\tablenum{1}
\tablehead{\colhead{$\langle z \rangle$} & \colhead{log SSFR} &
  \colhead{log SSFR} & \colhead{log SSFR}\\& \colhead{(1)} & \colhead{(2)} & \colhead{(3)}}
\startdata
0.60 & 0.40 $\pm$ 0.04 & $-$0.20 $\pm$ 0.03 & $-$1.11 $\pm$ 0.10 \\
1.00 & 0.40 $\pm$ 0.02 & $-$0.09 $\pm$ 0.03 & $-$1.04 $\pm$ 0.06 \\
1.50 & 0.41 $\pm$ 0.02 & $-$0.07 $\pm$ 0.02 & $-$1.02 $\pm$ 0.08 \\
2.15 & 0.59 $\pm$ 0.03 &    0.16 $\pm$ 0.03 & $-$0.56 $\pm$ 0.07 \\
3.00 & 0.62 $\pm$ 0.03 &    0.19 $\pm$ 0.02 & $-$0.21 $\pm$ 0.08 \\
4.25 & 0.75 $\pm$ 0.04 &    0.22 $\pm$ 0.04 & $-$0.20 $\pm$ 0.08\\[-0.75em]\enddata
\tablecomments{The SSFR is given in units of Gyr$^{-1}$. (1) $\log
  M_\star/M_\odot \in [8.5,9.5]$. (2) $\log
  M_\star/M_\odot \in [9.5,10.5]$. (3) $\log M_\star/M_\odot \in
  [10.5,11.5]$.}
\end{deluxetable}

This important finding is evident in Fig.~\ref{f:meanssfrz}, where we
show the average SSFR of galaxies with different masses as a function
of redshift.  While at redshifts $z \lesssim 2$ the most massive
galaxies are in a quiescent state, at redshifts $z \gtrsim 2$ the SSFR
for massive galaxies increases by a factor of $\sim 10$ reaching the
epoch of their formation at $z \sim 2$ and beyond. While there is
evidence for dry merging in the field galaxy population
\citep{Faber2005, Bell2005b}, this strong increase in the SSFR of the
most massive galaxies suggests that at least part of this population
was formed in a period of efficient star formation in massive haloes.

\acknowledgments

We thank the anonymous referee for his comments which helped to
improve this Letter. We acknowledge funding by the DFG (SFB 375). This
research has made use of NASA's ADS Abstract Service.

\bibliographystyle{apj}
\bibliography{apjmnemonic,literature}

\begin{thebibliography}{}

\bibitem[\protect\citeauthoryear{{Arnouts} et~al.}{{Arnouts}
  et~al.}{2001}]{Arnouts2001}
{Arnouts}, S., et~al. 2001, A\&A, 379, 740

\bibitem[\protect\citeauthoryear{{Bauer} et~al.}{{Bauer}
  et~al.}{2005}]{Bauer2005}
{Bauer}, A.~E., {Drory}, N., {Hill}, G.~J.,  \& {Feulner}, G. 2005, ApJ, 621,
  L89

\bibitem[\protect\citeauthoryear{{Bell} et~al.}{{Bell}
  et~al.}{2005a}]{Bell2005b}
{Bell}, E.~F.,  et~al. 2005a, ApJ, submitted, astro-ph/0506425

\bibitem[\protect\citeauthoryear{{Bell} et~al.}{{Bell}
  et~al.}{2005b}]{Bell2005}
{Bell}, E.~F., et~al. 2005b, ApJ, 625, 23

\bibitem[\protect\citeauthoryear{{Bender} et~al.}{{Bender}
  et~al.}{2001}]{photred}
{Bender}, R.,  et~al. 2001, in Deep Fields, ed. S.~{Cristiani}, A.~{Renzini},
  \& R.~E. {Williams} (Springer), 96

\bibitem[\protect\citeauthoryear{{Brinchmann} et~al.}{{Brinchmann}
  et~al.}{2004}]{Brinchmann2004}
{Brinchmann}, J., {Charlot}, S., {White}, S.~D.~M., {Tremonti}, C.,
  {Kauffmann}, G., {Heckman}, T.,  \& {Brinkmann}, J. 2004, MNRAS, 351, 1151

\bibitem[\protect\citeauthoryear{{Brinchmann} \& {Ellis}}{{Brinchmann} \&
  {Ellis}}{2000}]{BE00}
{Brinchmann}, J.,  \& {Ellis}, R.~S. 2000, ApJ, 536, L77

\bibitem[\protect\citeauthoryear{{Bruzual} \& {Charlot}}{{Bruzual} \&
  {Charlot}}{2003}]{BC2003}
{Bruzual}, G.,  \& {Charlot}, S. 2003, MNRAS, 344, 1000

\bibitem[\protect\citeauthoryear{{Calzetti}}{{Calzetti}}{1997}]{Calzetti1997}
{Calzetti}, D. 1997, AJ, 113, 162

\bibitem[\protect\citeauthoryear{{Casertano} et~al.}{{Casertano}
  et~al.}{2000}]{HDFSb}
{Casertano}, S., et~al. 2000, AJ, 120, 2747

\bibitem[\protect\citeauthoryear{{Cimatti} et~al.}{{Cimatti}
  et~al.}{2004}]{Cimatti2004}
{Cimatti}, A., et~al. 2004, Nature, 430, 184

\bibitem[\protect\citeauthoryear{{Cowie} et~al.}{{Cowie}
  et~al.}{1996}]{Cowie1996}
{Cowie}, L.~L., {Songaila}, A., {Hu}, E.~M.,  \& {Cohen}, J.~G. 1996, AJ, 112,
  839

\bibitem[\protect\citeauthoryear{{Drory} et~al.}{{Drory}
  et~al.}{2004}]{munics6}
{Drory}, N., {Bender}, R., {Feulner}, G., {Hopp}, U., {Maraston}, C.,
  {Snigula}, J.,  \& {Hill}, G.~J. 2004, ApJ, 608, 742

\bibitem[\protect\citeauthoryear{{Drory}, {Bender}, \& {Hopp}}{{Drory}
  et~al.}{2004}]{masscal}
{Drory}, N., {Bender}, R.,  \& {Hopp}, U. 2004, ApJ, 616, L103

\bibitem[\protect\citeauthoryear{{Drory} et~al.}{{Drory}
  et~al.}{2001}]{munics1}
{Drory}, N., {Feulner}, G., {Bender}, R., {Botzler}, C.~S., {Hopp}, U.,
  {Maraston}, C., {Mendes de Oliveira}, C.,  \& {Snigula}, J. {2001}, MNRAS,
  325, 550

\bibitem[\protect\citeauthoryear{{Drory} et~al.}{{Drory} et~al.}{2005}]{fdfmf}
{Drory}, N., {Salvato}, M., {Gabasch}, A., {Bender}, R., {Hopp}, U., {Feulner},
  G.,  \& {Pannella}, M. 2005, ApJ, 619, L131

\bibitem[\protect\citeauthoryear{{Faber} et~al.}{{Faber}
  et~al.}{2005}]{Faber2005}
{Faber}, S.~M.,  et~al. 2005, ApJ, submitted, astro-ph/0506044

\bibitem[\protect\citeauthoryear{{Feulner} et~al.}{{Feulner}
  et~al.}{2003}]{munics5}
{Feulner}, G., {Bender}, R., {Drory}, N., {Hopp}, U., {Snigula}, J.,  \&
  {Hill}, G.~J. 2003, MNRAS, 342, 605

\bibitem[\protect\citeauthoryear{{Feulner} et~al.}{{Feulner}
  et~al.}{2005}]{munics7}
{Feulner}, G., {Goranova}, Y., {Drory}, N., {Hopp}, U.,  \& {Bender}, R. 2005,
  MNRAS, 358, L1

\bibitem[\protect\citeauthoryear{{Fontana} et~al.}{{Fontana}
  et~al.}{2003}]{Fontana2003}
{Fontana}, A., et~al. 2003, ApJ, 594, L9

\bibitem[\protect\citeauthoryear{{Fontana} et~al.}{{Fontana}
  et~al.}{2004}]{Fontana2004}
{Fontana}, A., et~al. 2004, A\&A, 424, 23

\bibitem[\protect\citeauthoryear{{Gabasch} et~al.}{{Gabasch}
  et~al.}{2004a}]{fdflf1}
{Gabasch}, A., et~al. 2004a, A\&A, 421, 41

\bibitem[\protect\citeauthoryear{{Gabasch} et~al.}{{Gabasch}
  et~al.}{2004b}]{fdfsfr}
{Gabasch}, A., et~al. 2004b, ApJ, 616, L83

\bibitem[\protect\citeauthoryear{{Guzman} et~al.}{{Guzman}
  et~al.}{1997}]{Guzman1997}
{Guzman}, R., {Gallego}, J., {Koo}, D.~C., {Phillips}, A.~C., {Lowenthal},
  J.~D., {Faber}, S.~M., {Illingworth}, G.~D.,  \& {Vogt}, N.~P. 1997, ApJ,
  489, 559

\bibitem[\protect\citeauthoryear{{Hammer} et~al.}{{Hammer}
  et~al.}{2005}]{Hammer2005}
{Hammer}, F., {Flores}, H., {Elbaz}, D., {Zheng}, X.~Z., {Liang}, Y.~C.,  \&
  {Cesarsky}, C. 2005, A\&A, 430, 115

\bibitem[\protect\citeauthoryear{{Heidt} et~al.}{{Heidt} et~al.}{2003}]{FDF1}
{Heidt}, J., et~al. 2003, A\&A, 398, 49

\bibitem[\protect\citeauthoryear{{Juneau} et~al.}{{Juneau}
  et~al.}{2005}]{Juneau2005}
{Juneau}, S., et~al. 2005, ApJ, 619, L135

\bibitem[\protect\citeauthoryear{{Labb{\' e}} et~al.}{{Labb{\' e}}
  et~al.}{2005}]{Labbe2005}
{Labb{\' e}}, I., et~al. 2005, ApJ, 624, L81

\bibitem[\protect\citeauthoryear{{Longhetti} et~al.}{{Longhetti}
  et~al.}{2005}]{tesis4}
{Longhetti}, M., et~al. 2005, MNRAS, 361, 897

\bibitem[\protect\citeauthoryear{{Madau}, {Pozzetti}, \& {Dickinson}}{{Madau}
  et~al.}{1998}]{Madau1998}
{Madau}, P., {Pozzetti}, L.,  \& {Dickinson}, M. 1998, ApJ, 498, 106

\bibitem[\protect\citeauthoryear{{Noll} et~al.}{{Noll} et~al.}{2004}]{fdfspec}
{Noll}, S., et~al. 2004, A\&A, 418, 885

\bibitem[\protect\citeauthoryear{{P{\' e}rez-Gonz{\' a}lez} et~al.}{{P{\'
  e}rez-Gonz{\' a}lez} et~al.}{2005}]{Perez2005}
{P{\' e}rez-Gonz{\' a}lez}, P.,  et~al. 2005, ApJ, in press, astro-ph/0505101

\bibitem[\protect\citeauthoryear{{P{\' e}rez-Gonz{\' a}lez} et~al.}{{P{\'
  e}rez-Gonz{\' a}lez} et~al.}{2003}]{PerezGonzalez2003}
{P{\' e}rez-Gonz{\' a}lez}, P.~G., {Gil de Paz}, A., {Zamorano}, J., {Gallego},
  J., {Alonso-Herrero}, A.,  \& {Arag{\' o}n-Salamanca}, A. 2003, MNRAS, 338,
  525

\bibitem[\protect\citeauthoryear{{Salpeter}}{{Salpeter}}{1955}]{Salpe55}
{Salpeter}, E.~E. 1955, ApJ, 121, 161

\bibitem[\protect\citeauthoryear{{Saracco} et~al.}{{Saracco}
  et~al.}{2005}]{tesis3}
{Saracco}, P., et~al. 2005, MNRAS, 357, L40

\bibitem[\protect\citeauthoryear{{Saracco} et~al.}{{Saracco}
  et~al.}{2003}]{tesis1}
{Saracco}, P., et~al. 2003, A\&A, 398, 127

\bibitem[\protect\citeauthoryear{{Schirmer} et~al.}{{Schirmer}
  et~al.}{2003}]{Schirmer2003}
{Schirmer}, M., {Erben}, T., {Schneider}, P., {Pietrzynski}, G., {Gieren}, W.,
  {Carpano}, S., {Micol}, A.,  \& {Pierfederici}, F. 2003, A\&A, 407, 869

\bibitem[\protect\citeauthoryear{{Smail} et~al.}{{Smail}
  et~al.}{2002}]{Smail2002a}
{Smail}, I., {Ivison}, R.~J., {Blain}, A.~W.,  \& {Kneib}, J.-P. 2002, MNRAS,
  331, 495

\bibitem[\protect\citeauthoryear{{Williams} et~al.}{{Williams}
  et~al.}{1996}]{HDFN}
{Williams}, R.~E., et~al. 1996, AJ, 112, 1335

\end{thebibliography}



\end{document}